\documentstyle[preprint,aps
                           ]{revtex}

\def\ltap{\ \raisebox{-.5ex}{\rlap{$\sim$}} \raisebox{.4ex}{$<$}\ }

\begin{document}

\draft
\pagestyle{empty}                                      

{\tighten
\preprint{\vbox{\hbox{NTUTH-94-18}\hbox{August 1994}\hbox{}\hbox{}}}

\title{Singlet Charge $2/3$ Quark hiding the Top:\\
        Tevatron and LEP Implications}

\author{Wei-Shu Hou}
\address{Department of Physics, National Taiwan University, Taipei,
Taiwan 10764, R.O.C.\thanks{Permanent address.}}
\address{Theory Division, CERN, CH-1211,
Geneva 23, Switzerland}
\author{Hsuan-Cheng Huang}
\address{Department of Physics, National Taiwan University, Taipei,
Taiwan 10764, R.O.C.}


\maketitle
\begin{abstract}
If $c$ and $t$ quarks are strongly mixed with a
weak singlet charge $2/3$ quark,
$BR(t\to \ell\nu + X)$ could be suppressed via the $t\to cH^0$ mode,
thereby the top quark could still hide below $M_W$,
whereas the heavy quark signal observed at the Tevatron is
due to the dominantly singlet quark $Q$.
This may occur without affecting the small $m_c$ value.
Demanding $m_Q \simeq 175$ GeV and $m_t \ltap M_W$,
we find that $BR(t\to \ell\nu + X)$ cannot be too suppressed.
The heavy quark $Q$ decays via $W,\ H$, and $Z$ bosons.
The latter can lead to $b$-tagged $Z + 4$ jet events,
while the strong $c$--$Q$ mixing is reflected in sizable $Q\to sW$ fraction.
$Z\to t\bar c$ decay occurs at tree level and
may be at the $10^{-3}$ order,
leading to the signature of $Z\to \ell\nu b\bar c$, all isolated
and with large $p_T$, at $10^{-5}$ order.
\end{abstract}

\pacs{}
}

\pagestyle{plain}

\section{Introduction}

Recently, the CDF collaboration has reported \cite{CDF} some evidence
for the production of heavy quarks with mass of order 174 GeV
at the Tevatron. The most likely explanation is, of course, the
standard model (SM) top quark.
However, at present, in principle it is still possible \cite{Hou}
that the signal is due to some other heavy quark, whereas
the actual top quark is hiding below $M_W$.
This is because the top quark semileptonic branching ratio (BR)
has not yet been measured. If, for some reason,
$B_{s.l.} \equiv BR(t\to b\ell\nu) \ll {1\over 9}$,
the SM expected value, the top quark may have evaded
detection. This can arise basically only through scalar induced
interactions \cite{Hou}.

One such scenario was proposed \cite{MN} earlier
by Mukhopadhyaya and Nandi (MN).
Following a suggestion \cite{BH} by Barbieri and Hall (BH),
MN considered the existence of $SU(2)$ singlet
charge ${2\over 3}$ quark $Q$ alongside SM fermions.
Since GIM is broken, the mixing of $Q$ with up-type quarks
induce tree level flavor changing neutral couplings of
the SM Higgs boson.
If $m_{H^0} < m_t$, $t\to cH^0$ transitions
may trigger the aforementioned mechanism of suppressing $B_{s.l.}$.
In a subsequent paper, facing criticisms of ``naturalness" \cite{CL,Hou2}
($1/m_Q$ suppressions of heavy $Q$ effects),
MN retracted, and considered $t\to cH^0$ dominance
to be not very likely \cite{MN2}.
In this letter we study the precise conditions that
$t\to cH^0$ dominance can be realised.
We find that this requires $Q$ to be strongly mixed with
{\it both} charm and top, which can occur even with a small $m_c$
eigenvalue.
However, we find that although $t\to cH^0$ can be dominant,
it is unlikely to be overwhelmingly dominant.
Thus, $t\to bW^*$ should occur at reduced but still
substantial fraction. This offers hope that,
even if $m_t < M_W$, the top quark can be uncovered
at the Tevatron by a renewed study with existing data.

The heavy quark $Q$ can decay both via $W$ and $Z$ bosons \cite{BP},
hence it could be the heavy quark observed by CDF.
Although one could not explain the larger than expected
cross section for 174 GeV quarks, one could
plausibly account for the $b$-tagged $Z + $ 4 jets events \cite{CDF}.
Eventually, events with $ZZ + $ 2 jets should start to emerge with
increased luminosity \cite{AEH}.
Another point of great phenomenological interest is
$Z\to t\bar c$ decays, first stressed in this context by BH \cite{BH}.
These occur at tree level again because of GIM violation.
Although widely known, the possibility apparently
has not been studied with actual LEP data because
of the standard expectation of a very heavy top.
We estimate \cite{BH} that $Z\to t\bar c$ could occur
at the $10^{-3}$ level, but
{\it with $BR(t\to b\ell\nu)$ of order a few percent}.
This results in a signal branching ratio of
$Z\to \ell^+\nu b\bar c$ at the few $\times 10^{-5}$ level,
and each LEP experiment could have a few tens of events at present.
The background level can probably be managed,
and LEP experiments are strongly urged to conduct
such a search.

\section{Singlet Quark Induced Couplings}

Besides the standard $u$-type quarks $u_{iL}^0$, $u_{iR}^0$
($i = 1 - 3$), we add a left-right singlet
charge $2/3$ quark $u_{4L}^0$, $u_{4R}^0$.
The left-handed singlet field $u_{4L}^0$ can pair up with
the four right-handed fields to form gauge invariant singlet masses,
which we denote as $M_i^\prime$ and $M$ respectively.
The right-handed singlet field $u_{4R}^0$ introduces three
extra Yukawa couplings, resulting in off-diagonal masses
which we denote as $m_i^\prime$.
Thus, the $u$-type quark mass matrix is
\begin{equation}
   \mbox{\boldmath $M$} = \mbox{\boldmath $Y$} + \mbox{\boldmath $S$}
   = \left[   \begin{array}{rr}
            m & m^\prime \\
            M^\prime & M
               \end{array} \right],
\end{equation}
where
\begin{equation}
\mbox{\boldmath $Y$} = \left[ \begin{array}{rr}
             m  & m^\prime \\
             0  & 0
            \end{array} \right],\;\;\;
\mbox{\boldmath $S$} = \left[ \begin{array}{rr}
             0  & 0 \cr
             M^\prime & M
           \end{array} \right],
\end{equation}
are Yukawa and singlet masses.
\mbox{\boldmath $M$} is diagonalised by a biunitary transform,
\begin{equation}
   \mbox{\boldmath $\overline M$} =
          U^\dagger \mbox{\boldmath $M$} U^\prime =
            \mbox{diag}\, (\bar m_u,\ \bar m_c,\ \bar m_t,\ \bar M_Q),
\label{eq:Mdiag}
\end{equation}
where, departing from the notation of MN \cite{MN},
\begin{equation}
U^\dagger = \left[ \begin{array}{rr}
                    K & x^\dagger \\
                    y^\dagger & z^*
                   \end{array} \right],
\end{equation}
and $K$ is a $3\times 3$ matrix.
The Yukawa matrix \mbox{\boldmath $Y$}
is not simultaneously diagonalised,
\begin{equation}
   \mbox{\boldmath $\overline Y$} =
          U^\dagger \mbox{\boldmath $Y$} U^\prime =
           \mbox{\boldmath $\overline M$} -
          U^\dagger \mbox{\boldmath $S$} U^\prime,
\end{equation}
and the off-diagonal term controls FCNC $H^0$ and $Z^0$ couplings.
The apparent freedom due to the presence of $U^\prime$ rotation
matrix on right-handed fields led MN originally to
conclude that $tcH^0$ coupling could easily be rather large.
However,
from eq. (\ref{eq:Mdiag}),
simple algebra gives
\begin{equation}
  - U^\dagger \mbox{\boldmath $S$} U^\prime =
  - \left[ \begin{array}{rr}
              x^\dagger\, x\mbox{\boldmath $\overline m$}
                                                      & x^\dagger z M_Q  \\
              z^*\, x\mbox{\boldmath $\overline m$}   & z^* z M_Q
                   \end{array} \right],
\end{equation}
where {\boldmath $\overline m$} is the diagonal $3\times 3$
mass matrix (see eq. (\ref{eq:Mdiag})).
We see that no reference to $U^\prime$ is left, and the
off-diagonal couplings depend only on mass eigenvalues and
$Q$-related mixing elements of $U$ \cite{Hou2,MN2}.
The relevant flavor changing Higgs couplings are \cite{Hou2}
($i \neq i^\prime$)
\begin{eqnarray}
 &-& ( \bar m_i x_{i^\prime}^* x_i \; \bar u_{i^\prime L} u_{iR} +
       \bar m_{i^\prime} x_i^* x_{i^\prime} \; \bar u_{iL} u_{i^\prime R})
                                      \, {H\over v},   \cr
\label{eq:Hu}
 &-& ( m_i z^* x_i   \; \bar Q_L u_{iR} +
       m_Q x_i^* z   \; \bar u_{iL} Q_R) \, {H\over v}.
\label{eq:HQ}
\end{eqnarray}
The FCNC $Z$ couplings are \cite{MN}
\begin{eqnarray}
 && {g\over 2\cos\theta_W}\, x_{i^\prime}^* x_i \;
           \bar u_{i^\prime L} \gamma_\mu u_{iL}\, Z^\mu + h.c.,   \cr
\label{eq:Zu}
 && {g\over 2\cos\theta_W}\, x_{i}^* z \;
                     \bar u_{iL} \gamma_\mu Q_L\, Z^\mu + h.c.,
\label{eq:ZQ}
\end{eqnarray}
which is simply related to the Higgs couplings.
The charged current becomes
\begin{eqnarray}
 && {g\over \sqrt{2}} V_{ij} \;
                     \bar u_{iL} \gamma_\mu d_{jL}\, W^\mu + h.c.,   \cr
\label{eq:Wu}
 && {g\over \sqrt{2}} y^\prime_j \;
                     \bar Q_{L} \gamma_\mu d_{jL}\, W^\mu + h.c.,
\label{eq:WQ}
\end{eqnarray}
where
\begin{equation}
V\equiv K\, U^{(d)}, \ \ \ \ \ \
y^\prime_j \equiv y_{i}^*\, U_{ij}^{(d)}.
\label{eq:KMyp}
\end{equation}
The $3\times 3$ KM matrix $V$ is no longer unitary.
Both $V$ and $y^\prime$ depend on the $3\times 3$
left-handed down quark rotation matrix $U^{(d)}$.

\section{Details}

We wish to
explore the range of parameter space where $tcH$
coupling could be sizable.
To this end we make a special choice of basis
to focus on the problem.
First, we choose $u_R$ fields
such that $M^\prime = 0$ in \mbox{\boldmath $S$}.
Second, we choose $u_{iL}^0$, $i = 1-3$, such that
the matrix $m$ is diagonal, hence the
KM matrix largely comes from the down-type quark sector
(we have checked that it is not possible to generate
the observed KM matrix structure just by introducing $u$-type singlet
quarks).
Only the charged current is affected by the $d$-type quark sector,
the FCNC Higgs and $Z$ couplings depend only on
$x_i$ and $z$.

The $u$-type quark mass matrix is now in the form
\begin{equation}
   \mbox{\boldmath $M$}
   = \left[   \begin{array}{cccc}
            m_1 & 0 & 0 & \Delta_1 \\
            0 & m_2 & 0 & \Delta_2 \\
            0 & 0 & m_3 & \Delta_3 \\
            0 & 0 & 0 & M
               \end{array} \right].
\label{eq:ansatz}
\end{equation}
The relevant freedom introduced by the singlet quark $Q$
is parametrized as 3 new off-diagonal Yukawa terms,
plus the diagonal, gauge invariant Dirac mass $M$.
The parameters $x_i$, $y_i$ and $z$ can
be found by diagonalizing \mbox{\boldmath $MM$}$^\dagger$.
Without loss of generality, we set $\Delta_1 = 0$
so $u$ quark decouples from our discussion.

To illustrate the correlation between $\hat m_i\equiv m_i/M$
and $\hat\Delta_i \equiv \Delta_i/M$,
we set $\Delta_2 = 0$ and plot, in Fig. 1, the mass eigenvalues
$m_t/M$, $M_Q/M$ vs. $\hat m_3$ for different $\hat\Delta_3$ values.
Level repulsion is evident: $m_t < m_3$ and $M_Q > M$
for $m_3$, $\Delta_3 < M$.
For larger $m_3,\ \Delta_3$ values,
we adopt the convention that, if $x_t > 0.5$,
the heavier state is defined as the top quark.
Thus, Fig. 1 depicts both the mass eigenvalues and the label
for $t$ and $Q$.

We are more interested in the effect of $\Delta_2$.
With finite $\Delta_2$, but negligible $m_1$, $\Delta_1$, $m_2$,
the heavy mass eigenvalues are
\begin{equation}
m_t^2,\ M_Q^2 = {\Sigma^2 \mp
                  \sqrt{\Sigma^4 - 4m_3^2\left(M^2 + \Delta_2^2\right)}
                  \over 2},
\end{equation}
where $\Sigma^2 = M^2 + m_3^2 + \Delta_3^2 + \Delta_2^2$.
For sake of discussion, we consider the case where
$m_i,\ \Delta_i < M$ (top lighter).
Note that in Fig. 1 when $\hat \Delta_3$ is not too large,
the top mass eigenvalue is close to the diagonal term $m_3$.
This is a generic feature.
When other $\Delta$'s can be ignored and
$\hat\Delta_i$ is not too big,
the mass eigenvalue and mixing are roughly
\begin{equation}
\bar m_i^2 \sim {m_i^2 \over 1 + \hat\Delta_i^2}, \ \ \ \ \
x_i \sim \hat\Delta_i.
\end{equation}
These relations become affected only when there are
{\it two} $\hat \Delta_i$ values that are sizable, which follows
largely as a consequence of unitarity of the $4\times 4$
matrix $U$.
In Fig. 2 we plot $x_t$, $x_c$ as a function of $\hat \Delta_3$ for
$\hat m_3 = 0.7$ and $\hat \Delta_2 = 0,$ $0.5$, $1$, $1.5$.
Note the remarkable feature that $x_c$ is almost independent of $x_t$,
but $x_t$ is suppressed by large $x_c$ through unitarity.
The physical reason for this can be traced back to the
fact that $V_{cb} \sim 0.04$ is very small compared to 1,
and that $m_c$ is small.

Thus,
the eigenvalue $\bar m_c$ could be made small by
choosing a small value for $m_2$,
but this does not forbid $\Delta_i$ from being
sizable.
This is precisely {\it counter} the ``hierarchy principle" \cite{dB}
advocated in ref. \cite{Hou2}.
However,
given that $m_3$, $\Delta_3$ are large,
other than being a prejudice, there is really
no reason why $\Delta_2$ cannot be large,
since it is an independent parameter.
Of course, if $\Delta_2 \sim m_2$, then the conclusions of
ref. \cite{Hou2} would hold.

\section{Phenomenology}

Inspection of eq. (\ref{eq:Hu}) suggests that
$t_L \to c_R$ transitions are suppressed by $m_c/v$ \cite{Hou2},
but $t_R \to c_L$ transitions have the effective coupling
$m_t\, x_c^* x_t/v$.
Since $m_t/v$ is not small,
so long that $\vert x_c x_t\vert$ is not too suppressed,
the $t\to cH^0$ mode has good probability to be dominant
over $t\to bW^*$ \cite{MN}.
The necessary condition is therefore that
both $x_c$ and $x_t$ are sizable and neither are suppressed.
Hence, {\it Q, t and c all become rather arbitarily mixed}
although the charm mass is fixed by $m_2$.
Such an unusual situation is bound to have unusual consequences
beyond $t \to cH^0$ being sizable.

To illustrate the possibility of $t\to cH^0$ dominance,
{\it and} at same time account for CDF's apparent observation of
a heavy quark of mass 174 GeV, we demand that the
heavier quark (whether dominantly doublet or singlet) mass to be
pinned to the CDF value.
We then choose $\hat\Delta_2,\ \hat\Delta_3 = 0.7,\ 0.75$,
$M = 110$ GeV, vary $m_3$ (to get $m_t$, $M_Q$ etc.)
and plot, in Fig. 3, $BR(t\to cH^0)$
vs. $m_t$ (the physical mass) up to 90 GeV,
for $m_H = $ 50--75 GeV.
We allow for $m_H$ below the present LEP bound
in case there are more than one Higgs doublet \cite{PDG}.
In producing Fig. 3, we compute the $t\to cH^0$ and $bW^*$ decay width
using the couplings of eqs. (\ref{eq:Hu}) and (\ref{eq:Wu}).
We assume that $U^{(d)}$ amounts to a ``small" rotation close to the
``standard" KM matrix,
ignoring all phases.
We have also ignored $t\to cZ^*$ decay as this is
a three body process subdominant compared to $t\to bW^*$.
It is clear that, if the Higgs boson mass is
sufficiently light, $t\to cH^0$ can be dominant.
However, the combined demand of $m_Q \simeq$ 174 GeV,
and $m_t \ltap M_W$,
dictates that the $t\to cH^0$ mode cannot
be overwhelmingly dominant. Thus, although suppressed,
$BR(t\to bW^*)$ should not be vanishingly small.
For larger $m_H$, $t\to cH^0$ dominance quickly fades,
and the possibility is ruled out by CDF since $t\to bW^*$ is
not drastically suppressed.
In the following, we shall assume that one works in the
domain where $B_{s.l.}$ for the ``top" (it could be the dominantly
singlet quark, since we do not know the scale for $M$)
is suppressed by $1/3$ or more. That is, $B_{s.l.} < 1/27$.

Fig. 3 corresponds to $x_c,$ $x_t \simeq 0.59,\ 0.53$, 
with $m_t,\ M_Q = 75,\ 174$ GeV.
The corresponding $Q\to bW,\ sW,\ tH,\ cH,\ tZ,\ cZ$ branching ratios
are 0.51, 0.24, 0.04, 0.1, 0.02, 0.09, respectively.
Note that, as a consequence of large $x_c$, $Q \to sW$ decay has
a sizable rate! The modes $Q\to tH^0$ and $tZ$ are suppressed
by phase space,
while $Q\to cH^0$ and $cZ$ are suppressed by an extra power of
$\vert x_c\vert^2$.
Thus, $W$ induced decays are still dominant, but the $b$ content in
final state is diluted slightly by the $Q\to sW$ mode.
Although one cannot account for the large production cross section
for the heavy quark
({\it one could always add another singlet $u$-type quark for this purpose}),
other features reported by CDF
can be accounted for \cite{BP,BMPS},
in particular, the appearance of $b$-tagged $Z+$ 4 jet events.
The $Z$ boson comes from $Q\to cZ$, $tZ$,
while a $b$-tag could come from $Q\to bW$,
or from $t\to bW^*$ or $H\to b\bar b$, etc.,
in subsequent decays.
This could be at $20\%$ of the $Q\bar Q$ cross section,
hence consistent to what is observed.
On the other hand, single-$W$ with $b$-tag is slightly depressed
($\sim 70\%$)
compared to the standard top.
We therefore conclude that the heavy quark observed by CDF
may well be a doublet-singlet mixed state $Q$.
The scenario offers many signatures
and can be checked experimentally.
The light top with not too suppressed $B_{s.l.}$ can perhaps be
probed with existing Tevatron data \cite{Hou}.

The scenario has a consequence that may be studied at LEP.
As first pointed out by Barbieri and Hall \cite{BH},
$Z\to t\bar c$ can be quite sizable with
existence of charge 2/3 singlet quarks.
Using eq. (\ref{eq:Zu}) and $x_c$, $x_t$ values for the example above,
we estimate that $BR(Z\to t\bar c + \bar t c)$ is of order
a few $\times 10^{-3}$, which is consistent with ref. \cite{BH}.
Other phenomenological constraints are not particularly stringent,
and can be found in ref. \cite{BH}.
For example, $D^0$--$\bar D^0$ mixing constraint can be satisfied
with small $\Delta_1$.
Since $BR(t\to \ell\nu X)$ does not vanish,
we estimate that the potentially observable signal of
$Z\to t\bar c \to \ell\nu + $ 2 jet
(where the jets contain $b$ and $c$)
could have a branching ratio of order a few $\times 10^{-5}$.
Since the lepton and neutrino should be well isolated with
sizable ($15 - 20$ GeV) $p_T$ or missing energy (they are {\it bona fide}
virtual $W$ decay events!), and that $\ell$, $\nu$ and one jet
should pair up to be the top mass,
there should be sufficient handles for the suppression of background.
The latter presumably comes from events with
$Z\to b\bar b$ plus gluon bremsstrahlung.


\acknowledgements
This work is supported in part
by the National Science Council of
the Republic of China
under grant NSC-83-0208-M-002-023.
WSH wishes to thank the CERN Theory Division for
hospitality and a stimulating environment,
and T. Ferguson for discussion.

\begin{figure}
\caption{$m_t/M$, $M_Q/M$ vs. $\hat m_3$ for various $\hat\Delta_3$ values.
         The solid(dash) lines stand for physical t(Q) quark.}
\label{fig1}
\end{figure}

\begin{figure}
\caption{Mixing parameter $x_t$, $x_c$
         vs. $\hat \Delta_3$ for $\hat m_3 = 0.7$ and
         $\hat \Delta_2 = $ 0(dots), 0.5(solid), 1(dash), 1.5(dotdash).
         Note that $x_c$ is basically independent of $x_t$.}
\label{fig2}
\end{figure}

\begin{figure}
\caption{$BR(t\to cH^0)$ vs. $m_t$
for $\hat \Delta_2,\ \hat \Delta_3 = 0.7,\ 0.75$,
$M = 110$ GeV,
and $m_H = $ 50--75 GeV in 5 GeV intervals.
}
\label{fig3}
\end{figure}

\end{document}